# Homologous self-assembled superlattices: What causes their periodic polarity switching? Review, model, and experimental test

Varun Thakur,[1] Dor Benafsha,[1] Yury Turkulets,[1] Almog R. Azulay,[1] Xin Liang,[2] Rachel S. Goldman,[3] and Ilan Shalish,[1]*

[1]School of Electrical Engineering, Ben-Gurion University, Beer Sheva 8410501, Israel.
[2] Beijing Institute of Nanoenergy and Nanosystems, Chinese Academy of Sciences, Beijing 101400, China, and School of Nanoscience and Technology, University of Chinese Academy of Sciences, Beijing 100049, China
[3]Department of Materials Science and Engineering, University of Michigan, 2300 Hayward, St., Ann Arbor, 48109, MI, USA

Quantum semiconductor structures are commonly achieved by bandgap engineering that relies on the ability to switch from one semiconductor to another during their growth. Growth of a superlattice is typically demanding technologically. In contrast, accumulated evidence points to a tendency among a certain class of multiple-cation binary oxides to self-assemble spontaneously as superlattice structures. This class has been dubbed the homologous superlattices. For a famous example, when a mixture of indium and zinc is oxidized, the phases of In-O and ZnO separate in an orderly periodic manner, along the ZnO polar axis, with polarity inversion taking place between consecutive ZnO sections. As we review here, the same structure has been observed when the indium was replaced with other metals, and perhaps even in ZnO alone. This peculiar self-assembled structure has been attracting research over the past decade. The purpose of this study is to gain understanding of the physics underlying the formation of this unique structure. Here, we first provide an extensive review of the accumulated literature on these spontaneously-formed structures and then propose an explanation for the long-standing mystery of this intriguing self-assembly in the form of an electrostatic growth phenomenon and test the proposed model on experimental data.

## I. INTRODUCTION

Much of the work on transparent conductive oxide semiconductor thin films and nanostructured devices has focused on zinc oxide [1,2,3,4,5] or ZnO-based compounds, namely In-Ga-Zn-O, Zn-Sn-O and In-Zn-O.[6,7,8] The general consensus behind choosing ZnO-based structures seems to be that for obtaining transparent materials with high conductivities, a class of oxides with multiple cation species with the electronic configuration of $(n-1)d^{10}s^0$ holds the most promise.[9] The empty metal cation s-orbitals generally form the conduction band in these materials, while the overlap between them gives the n-conduction pathway, making their electronic configuration most suitable for various device applications.

Among the ZnO-based semiconductor-oxides, In-Zn-O has received special attention for an interesting phenomenon. Thin films of In-Zn-O were observed to possess a superlattice structure following the relation $In_2O_3(ZnO)_m$ and were therefore considered to be a homologous series.[10] This unique structure holds a special interest due to the spatial confinement of conduction electrons in 2-dimensional layers, affording them exceptional electron transport properties in addition to their being excellent transparent oxide semiconductors materials.[11,12] In 2003-04, Nomura *at al.* demonstrated transparent field effect transistors fabricated on a complex $InGaO_3(ZnO)_5$ superlattice structure which showed excellent properties including a carrier mobility of 80 cm$^2$V$^{-1}$s$^{-1}$, turn on voltage of -0.5 V, and an on-off ratio of $10^6$[13,14,15]. Following this demonstration, other groups became engaged in fabrication of transistors using In-Zn-O [16,17,18,19] and Sn-Zn-O [20,21], In-Ga-Zn-O,[22,23,24,25,26] and Al-Zn-O.[27] Field effect mobilities ranging from 5-50 cm$^2$V$^{-1}$s$^{-1}$ depending on annealing temperatures were reported with turn on voltages between -5 and 15 V.[20] Devices fabricated by Dehuff *at al.* [16] also showed excellent properties with mobilities reaching 55 cm$^2$V$^{-1}$s$^{-1}$ and turn on voltages ~ -20 V. Research has also been carried out on applications such as photocatalysis.[28] These papers mark the onset of intense research of these self-assembled ZnO-based superlattice structures for various electronic, optoelectronic, and thermoelectric applications. In the following years, several papers reported materials possessing the general electronic configuration of $(n-1)d^{10}s^0$ (n≥4), among which the most common superlattice structures were those observed in Zn-Sn-O, In-Zn-O and Ga-In-Zn-O.

In this paper, we attempt to explain the physics underlying the spontaneous formation of the so-called homologous compound superlattice structures. To this end, we first review the published literature. While there have already been published reviews mentioning the superlattice structure, they have not focused on the mechanism of its formation.[29] Here, we follow our literature review with a proposed model and its experimental verification. In the most commonly studied of these materials, In-Zn-O, the ZnO flips its polarity periodically along the polar axis, while the indium may be viewed as a *polarization-inversion promoter*. Understanding of the causes for the formation of periodic polarity switching is currently lacking as well. We propose a simple electrostatic model that explains the phenomenon as a periodic charging process taking place during the crystal growth, due to the strong polar nature of the main material. Our model and results emphasize the important role of electrostatics in the growth of polar semiconductors – a mechanism of which the semiconductor material community has mostly been unaware.

## II. REVIEW

Already at an early stage, In-Zn-O compound crystals were identified to occur at specific sequences of layers that led to their classification as inorganic homologous series. Sections of m monolayers of ZnO were found to be periodically sandwiched between monolayers of the structure $InMO_3$, following the formula $InMO_3(ZnO)_m$ (where M = In, Fe, Ga, Al, m = integer). Interestingly, the layers self-assemble to form this ordered periodic sequence of alternating phases and do so both in films, and in nanowires. While there have been numerous reports on the synthesis and characterization of these structures, a comprehensive review is lacking so far. In this paper, we propose a model to explain the physics underlying the formation of this complex structure. We propose a mechanism that explains why this self-assembly process takes place. To introduce the reader to the problem, we first review the superlattice structures of the homologous $InMO_3(ZnO)_m$ series with a focus on $In_2O_3(ZnO)_m$, which was one of the earliest of these compounds to have been synthesized and studied, and has also been the most commonly reported among them.

### 2.1 Early work on bulk and thin films

Initial observations of these structured materials were made in *ceramic polycrystalline thin films*. Synthesis of what was thought to be $In_2O_3(ZnO)_m$ (m = 2-5 and 7) having a layered wurtzite structure was first reported in 1967 by Kasper.[30] However, two decades elapsed before the first attempt by Cannard and Tilley in 1988 to analyze the structure using high resolution transmission electron microscopy (HRTEM).[31] They reported the presence of metal-oxygen layer stacks perpendicular to the c-axis in the hexagonal crystal system. Their seminal study gave an important insight into the actual atomic order in this superlattice. Nakamura *at al.* then synthesized and analyzed the phase relations in the $In_2O_3$-$ZnGa_2O_4$-ZnO system at 1350°C and proposed homologous phases having solid solutions of $(InGaO_3)_2ZnO$, $In_{1.33}Ga_{0.67}O_3(ZnO)$-$InGaO_3(ZnO)$-$In_{0.92}Ga_{1.08}O_3(ZnO)$, $In_{1.68}Ga_{0.32}O_3(ZnO)_2$-$InGaO_3(ZnO)_2$-$In_{0.68}Ga_{1.32}O_3(ZnO)_2$,

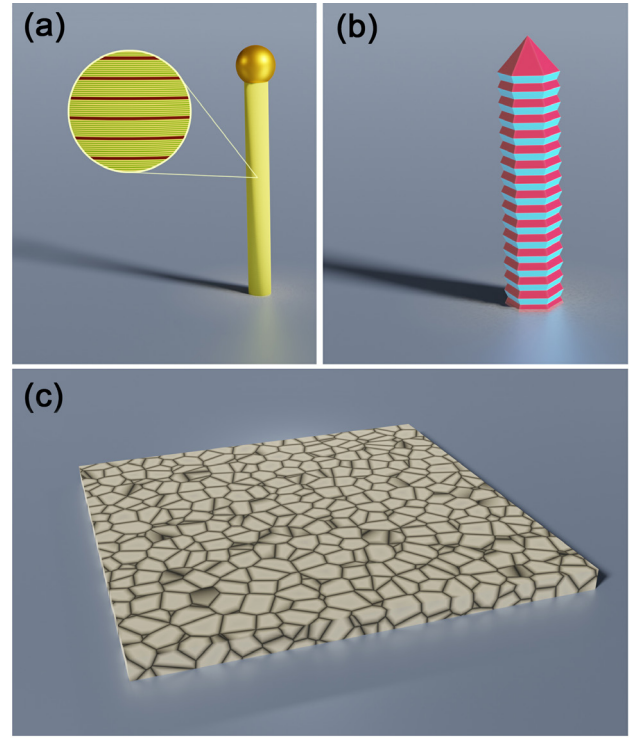

**Fig.** 1 The ZnO-based homologous crystal structures are commonly observed as (a) nanowires, (b) nanorods, and (c), polycrystalline layers. They are easily identifiable in nanorods by the corrugated surface caused by the periodically inverted domain layers. However, this surface structure is typically missing in nanowires. This is probably for the same reason that crystal facets are only observed in thick enough wires.

and $In_2O_3(ZnO)_m$-$InGaO_3(ZnO)_m$ -$In_{1-x}Ga_{1+x}O_3(ZnO)_m$ (m = 3-13)(0<x<1).[32] Lattice constants of the solid solutions of the homologous phases were measured using powder X-ray diffraction assuming the $InGaO_3(ZnO)_m$ system to be isostructural to the $LuFeO_3(ZnO)_m$ family without any single crystal data. The $ZnO$-$ZnGa_2O_4$ system showed no binary compounds. Rather, they claimed to have detected a solid solution of the ZnO phase, $(Ga_2O_3)_x(ZnO)_{1-x}$ ($0 \leq x \leq 0.093$). As expected, the ZnO phase (x = 0) showed a wurtzite structure, but the assumed solid solution that included the $Ga_2O_3$ compound appeared to deviate from the wurtzite structure to one with a lower symmetry as the value of x increased. Moriga *at al.* reported equilibrium phase relationships within the same temperature range of 1100°C-1400°C and also explored the physical properties of the stable phases.[33] They proposed a phase diagram that disagreed with the results of Nakamura *at al.* who reported stable phases of $Zn_8In_2O_{11}$ and possibly $Zn_{10}In_2O_{13}$ at 1350°C. Overall, they were able to observe phases with *k* values ranging from 3 to 15 over their temperature range, while the stability of their observed compounds increased with the temperature. Room temperature conductivity was found to increase as *k* decreased and was attributed to increased carrier concentration as well as higher mobility. Further experiments extended the reported trend to lower *k* values as well.[34] Sub-solidus phase relationships in the ZnO-In$_2$O$_3$-SnO$_2$ system were investigated by Harvey *at al.*[35] They confirmed the findings of Moriga *at al.* and were able to detect all the seven compounds in their phase space. Differing methods of preparation prompted the question as to whether a microstructural change was causing the stability and the change in physical properties. The experiments of Moriga, and their consequent studies of pulsed laser deposited films, posed an interesting conundrum. Dupont *at al.* reported microstructural studies of films made using both techniques while encompassing the entire composition range.[36] They used X-ray diffraction measurements and followed them with HRTEM imaging and diffraction. Their results gave a detailed microstructural database of the entire compositional range, while the method of preparation was not found to be a significant factor in the quality of the obtained films. Following this, Isobe reported single crystal diffraction data of LuFeO$_3$(ZnO)$_m$ (m = 1, 4-6).[37] This was the first of a series of studies of the RMO$_3$(M'O)$_m$ systems (R = In, Sc, Y, or a rare earth element; M = Fe, Ga, Al; M' = divalent cation, m = integer). They found that Lu occupied the octahedral site, while both Fe and Zn were present in trigonal bipyramidal sites. The LuFeO$_3$(ZnO)$_m$ crystal is composed of LuO$^{2-}$ and (FeZn$_m$)O$^+_{m+1}$ layers that alternate along the polar c-axis in the wurtzite system. Adding on to this work, the group reported single crystal data of the homologous series In$_2$O$_3$(ZnO)$_m$ (m = 3, 4 and 5), InGaO$_3$(ZnO)$_3$, and Ga$_2$O$_3$(ZnO)$_m$ (m = 7, 8, 9 and 16) in the In$_2$O$_3$-ZnGa$_2$O$_4$-ZnO system.[38] Their report of the In$_2$O$_3$(ZnO)$_m$ system had two main observations: (i) Both In$_2$O$_3$(ZnO)$_m$ and InGaO$_3$(ZnO)$_3$ were isostructural with LuFeO$_3$(ZnO)$_m$ and had a space group R$\bar{3}$m for m = odd or P6$_3$/mmc for m = even. (ii) Crystal structure models for In$_2$O$_3$(ZnO)$_m$ and InGaO$_3$(ZnO)$_m$ estimated from powder X-ray diffraction were consistent with the single crystal data.

Studies of the crystalline and electronic structure were still scarce, and most of the reports detailed phase relations. The data on single crystal experiments were limited at that stage to Weissenberg photographs. Even towards the end of the 1990s, as HRTEM use was becoming more common, much of the reported work was still based on x-ray diffraction. Schinzer *at al.* reported a study using Rietveld refinement and band structure calculations, leading to deeper insight into the crystal chemistry.[39] Taking the structure Zn$_3$In$_2$O$_6$ as a case study, the experimental part was carried out by synthesis and X-ray diffraction measurement of the sample. Rietveld refinement on the data was carried out, trying out the parameters in the two prevailing models of Kasper and Kimizuka. It was found that the Kasper model did not work well, while the Kimizuka model was more successful. The structural model suggested by Kasper bases the 'a' lattice parameter of Zn$_3$In$_2$O$_6$ on the zincite structure, and this way, a part of the structure can indeed be explained by similarities with ZnO. However, the In containing layer was explained as a purely bixbyite-type In$_2$O$_3$ structure, because of a similar behavior observed in Yb$^{3+}$ and Lu$^{3+}$, both of which having similar chemistry as In$^{3+}$ and crystallize in the bixbyite phase,

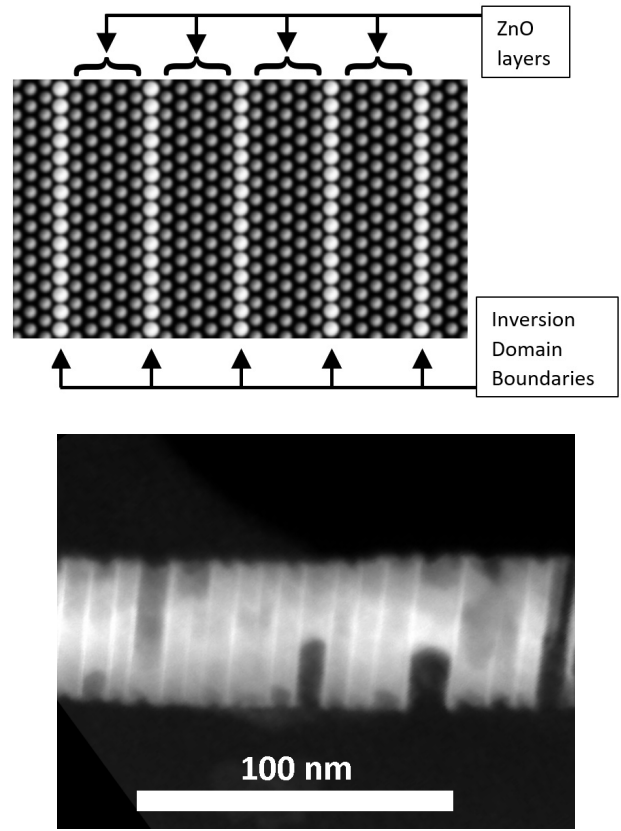

**Fig. 2** Top - An illustration showing the typical structure revealed in transmission electron microscope image: slabs of several (m) ZnO layers (here for example m=4) are separated by monolayers of another metal oxide. The polarity of each ZnO slab is inverted relative to those of its neighboring slabs, so that the other metal oxide monolayers form inversion domain boundaries. Images of similar structures have been shown, for example, by Ohta *at al.*, [11] Guo *at al.*,[76] Park *at al.*[93] and by Cao *at al.*,[120] Bottom – TEM image of In-Zn-O nanowire. The bright lines traversing the wire mark the In-O inversion domain boundaries.

and this model did not have a strong foundation. The Kimizuka model includes three different types of layers applied to mixed oxides in the MM'O$_3$(ZnO)$_m$ type structure, where M = Yb, Lu or In and M' being a trivalent metal cation. The main difference between the two models is in the varying stoichiometry in the layers, MO$_{1.5}$, M'ZnO$_{2.5}$ and ZnO. As opposed to earlier observations, Schinzer *at al.* did find similarities between Zn$_3$In$_2$O$_6$, and LuFeO$_3$(ZnO)$_m$, but observed a complete disorder of In and Zn in the MO layers. They proposed that the structure was layered having (Zn/In)O layers built from trigonal bipyramids stacked between sheets of InO$_2$. In the second part of their paper Schinzer *at al.* report band structure density functional theory (DFT) calculations of Zn$_3$In$_2$O$_6$, which had not been reported for these materials previously. For comparison they also calculate In$_2$O$_3$. Comparing with the density of states of In$_2$O$_3$ which is also previously unreported, the authors observe similar features such as high dispersion of lower conduction band and O (2s) bands at VBM, even when the crystal structure of Zn$_3$In$_2$O$_6$ is considered to be cubic or rhombohedral. In conclusion, while they find the model of

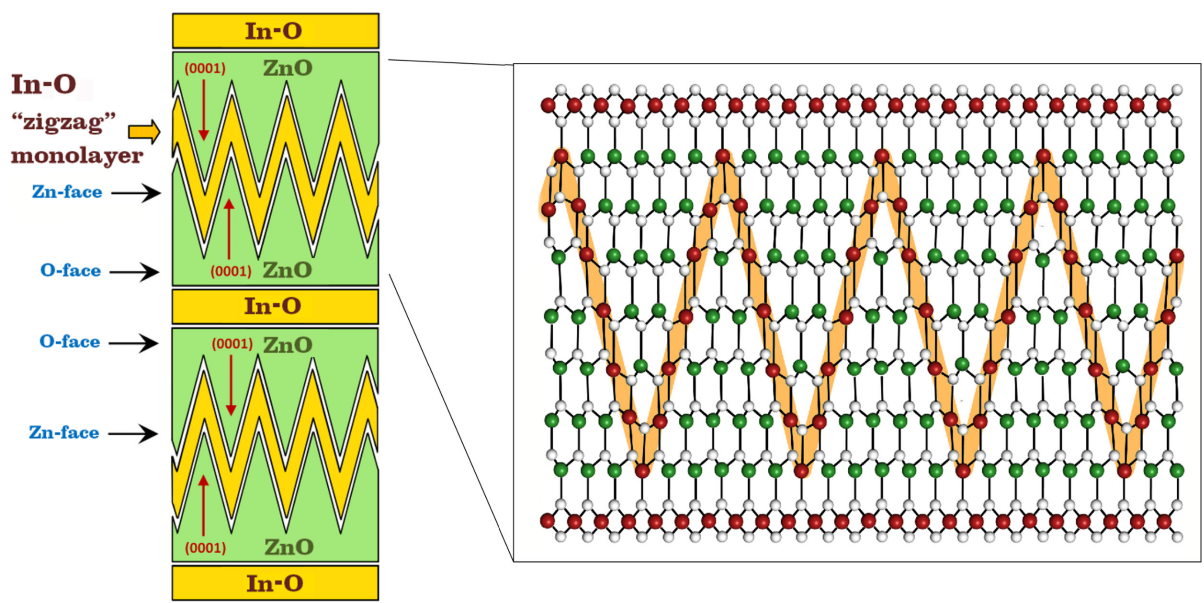

**Fig. 3** Illustration of the In-Zn-O superlattice structure, where each ZnO slab has an opposite polarity to its neighboring slabs, and the boundaries between the slabs are flat on the O-face and zigzagged on the Zn-face. Three decades elapsed before the zigzag boundary structure was realized. During this time the two ZnO slabs neighboring the zigzag layer were thought to be a single In-Zn-O ternary layer. Part of the difficulty may be related to the fact that the zigzag boundary are not always present and sometimes are replaced with an planar In-O layer.

Kimizuka *at al.* to be closer to their results and the conformation of $InO_2$ similar to that of $LuFeO_3(ZnO)_m$ family, but as opposed to the Kimizuka model, they find the cationic ordering to be more statistically distributed than allocated. Similarities between band structures of $In_2O_3$ and $Zn_3In_2O_6$ were found irrespective of the profound differences among the crystal structures that were compared, and a weak influence of filled *d* states of metal ions on the band structures was found near the Fermi level.

Up to that point, it was established that the $In_2O_3(ZnO)_m$ system followed the crystal structure pattern of $LuFeO_3(ZnO)_m$ and hence should have $InO^{2-}$ and $(ZnO)_m$ layers alternating along the c-axis of the wurtzite crystal system (Fig. 2). To confirm this finding visually, several groups carried out HRTEM imaging with electron diffraction.

Using HRTEM, Cannard and Tilley had showed that $In_2O_3(ZnO)_m$ (m = 4-7, 9 and 11) consisted of $In_2O_3$ layers interleaved with m layers of ZnO. However, Uchida *at al.* found evidence of a modulated structure in the homologous series $InFeO_3(ZnO)_m$ (m = 1, 6 and 13) [40] and in HRTEM on $In_2O_3(ZnO)_m$ structures (m = 3, 6, 10, 13, 15, 17 and 20), they indeed observed a modulated structure for m>6.[10] A lattice image of this modulated structure revealed a wavy contrast forming a *zig-zag shape*. Non-integral modulation periodicity was found to be linearly proportional to increasing *m* numbers. Based on lattice images of $In_2O_3(ZnO)_3$, a structural model was constructed consisting of $InO^{2-}$ (In-O) layers interleaved with four layers of In/Zn-O. The In atoms in the $InO^{2-}$ layers occupied octahedral sites, while the In and Zn atoms in the In/Zn-O layers occupied tetrahedral and trigonal bipyramidal (tbp) sites. Also, the ratio of the metals In and Zn in the In/Zn-O was 1:3 and they were randomly distributed in the metal sites. Hence, according to this model, the structure along the c-axis is composed of three In-O layers and twelve In/Zn-O layers per unit cell. Electron diffraction patterns in their report show two kinds of spots, one heavy, assigned to the In atoms residing in the octahedral sites in the In-O layer and the other gray, ascribed to both the In and Zn atoms in the In/Zn-O layer. The structure is calculated for a crystal thickness of 2 nm along the c-axis. For larger values of *m* (6 and 13), the structure modifies as observed from the diffraction patterns. The $In_2O_3(ZnO)_6$ structure shows the same brightness pattern as for m = 3, with In atoms in the In-O layers being heavy and both In/Zn atoms in the In/Zn-O layers being gray. However, one half of the unit cell now consists of 1 In-O layer interleaved with 7 In/Zn-O layers with a periodicity of 2.2 nm along the c-axis. The structure with m = 13 clearly shows the wavy patterns in the lattice image, and the corresponding diffraction pattern also reveals the presence of a few weak spots, indicating the modulated structure. The periodicity of this modulated structure along (110) is non-integral with a value of 4.9 nm. The lattice image showed the wavy patterns in a zig-zag fashion in the In/Zn-O layers. The periodicity of this modulation was observed to be inconsistent, varying from 28 lattice planes (4.6 nm) to about 31 lattice planes (5.1 nm). The angle of the zig-zag to the normal to the c-plane (normal of the c-axis) was calculated to be around 58°. Increasing the value of *m* further to 20, the same modulation is observed, with a more clearly defined zig-zag structure retaining the 58° angle. The non-integral periodicity also increases to about 7.1 nm along (110). The diffraction pattern recorded for these structures shows streaks along c-direction accompanying spots, indicating intergrowth structures with different m values. The image contrast in the modulations shown as the zig-zag or triangular pattern is also enhanced at the regions where the crystal is thicker, leading to the idea that the contrast could be due to the lattice distortion of $ZnO_4$ tetrahedron because of the local ordering of In atoms in the In/Zn-O layer. A superspace group based approach to describe the atomic arrangements as observed from HRTEM was also undertaken by the authors in a separate report.[41] It was shown that subcell structures in compounds with odd and even values of m can be described by monoclinic and orthorhombic cells, respectively. These experiments of the Kimizuka group hence showed the presence of a modulated, zig-zag, structure, which was absent in the results of Cannard and Tilley, because of the higher *m* values used here.

The finding of the zig-zag structure was an important step towards the understanding of the structure, because it previously led to the wrong conclusion that the In was part of a ternary In-Zn-O phase, whereas now, it became clear that some of the previously assumed ternary phases were actually two inverse-polarity ZnO domains with In-O decorated zigzag boundary.

A combination of simulation and experimental study was carried out by McCoy *at al.*[42] to verify the correct structural model by considering the defect energies of In atoms within the ZnO lattice. Modeling the inclusion of $In_2O_3$ layers as isolated point defects, defect clusters, and as an intergrowth of extended planar defects periodically arranged within the lattice, the authors compare their simulations with HRTEM. In the simulations, previously reported structures by Kasper, Kimizuka and Cannard-Tilly were also studied, among which the Cannard-Tilly model showed the least lattice energy. This means that the structure with a double indium oxide layer (bixbyite) separating slabs of wurtzite ZnO was found to be the most energetically favorable among the structures reported so far. Experimental validation for the studied models is slightly challenging since similarities in the space groups of different structures make it impossible to identify the planar defect structure solely on the basis of electron diffraction or X-ray methods. The HRTEM technique is able to distinguish between the various defect structures in a more unambiguous way and is hence used as a primary experimental tool for the same. Simulating HRTEM images with experimental parameters such as spherical aberration and beam divergence close to the actual setup showed that the Kasper model had a poor match with the experimental data while the Cannard-Tilly and Kimizuka models gave a much better match. Among these two, a closer inspection revealed subtle differences due to the structure of $InO_2$ layer and a projection of the simulation with the experimental data finally showed the best match between the Cannard Tilly model when considering the $Zn_4In_2O_7$ structure viewed along the $(1\bar{2}10)$ projection. The authors conclude that simulations show a stabilization of indium incorporation in the ZnO lattice by the formation of an indium rich planar defect oriented along the basal (0001) plane of the ZnO wurtzite structure. Finally, their proposed model is based on a wurtzite ZnO structure consisting of two indium layers along the (0001) direction separated by m ZnO wurtzite layers; each wurtzite region displaced from the next by a translation of $1/3(10\bar{1}0)$.

Based on Z-contrast scanning tunneling electron microscopy imaging, Yan *at al.* suggested the presence of a polytypoid structure.[43] Their results provided a crucial insight into the structure of $In_2O_3(ZnO)_m$ films, where irrespective of the *m* value, the number of ZnO layers between each In-O layer was found to be inconsistent. In other words, *the structure along the c-axis had variable periodicity*. Further, the analysis of the Z-contrast image to ascertain the complete structure also revealed that the In-O layer containing O atoms in an octahedral position form a wurtzite structure with the Zn atoms below, necessitating an *inversion of polarity in the ZnO slab following the In-O layer*. This was also the first report of the In-O layers forming an inversion domain boundary (IDB) in the $In_2O_3(ZnO)_m$ structure. Since each ZnO slab is bound by two In-O layers, the ZnO layer itself must contain another domain boundary. This other domain boundary was named by the authors a "mirror domain boundary". In that study, this domain boundary was *not* observed to be zigzagged.

The finding of periodically alternating polarities was a major leap in the understanding of the structure. This realization was delayed for a long time due to the inconsistent occurrence of the zigzag shape at one of the domain inversion boundaries (Fig. 3). As long as this zigzag boundary had not been realized to be a boundary, it was thought that the material between two flat boundaries was a solid solution InZnO. We note here that the structure closely resembles that of periodically twinned ZnO. However, twinning is only one of three types of this boundary, with the other two types being the flat mirror boundary (not twinning) and the zigzag boundary, which structure has yet to be fully understood. The decoration of the inversion domain boundaries with the other metal has also been misleading as it does not always take place in the same manner. In the mirror boundary they just replace Zn, while in the twin boundary they form an In-O structure. Several theoretical studies following the discovery of the zigzag boundary were undertaken to explain why this modulated boundary was energetically favorable.

## 2.2 Observations in Nanowires

As of the year 2000, and following the buildup of interest in nanowires, structural analysis was made easy. Nanowires do not require the laborious preparation of TEM cross-section layers. Their structure itself provides a ready-made TEM cross-section sample. No small advantage was the fact that at no additional complexity, one could easily grow a superlattice nanowire at a time when much effort was invested into adding various functionalities to the new "nanotechnology building block". From this point and on, most of the studies were carried out on nanowires.

The first report of $In_2O_3(ZnO)_m$ nanowires was by Jie *at al.*[44] They observed the formation of the superlattice structure only in a part of the substrate area (20-40%). HRTEM images and corresponding selective area electron diffraction (SAED) patterns revealed the presence of a superlattice, reflected as a series of smaller diffraction spots between two adjacent main spots (commonly dabbed "streaking"). The authors proposed an equation to relate the m values with 'd', with the distance between two adjacent In-O layers given by:

$$d[\text{Å}] = 6.349 + 2.602 \cdot m \qquad (1)$$

Using high resolution images, *d* was calculated to be about 53 Å, giving an *m* value of ≈18, and the average superlattice structure was concluded to be $In_2O_3(ZnO)_{18}$. The zig-zag modulation was observed here as well, and the authors attribute it to relief of stress caused by the introduction of the larger $In^{3+}$ ions into the ZnO lattice. An important deviation from the modulations observed in ceramic thin films was the fact that apart from being non-uniformly distributed among the In/Zn-O layers, the angles made by the zig-zag patterns with the normal to the c-axis ranged in these nanowires from 30° to 60°. As can be recalled from the previous discussion, the thin film samples always showed a consistent periodicity and a constant angle of 58°. The reason for this angle inconsistency in nanowires was suggested by the authors to be a result of the nanowire synthesis temperature (at the range of 800-1000°C) which is significantly lower in comparison to the film growth temperatures (greater than 1200°C). HRTEM images showed the lattice image of a nanowire with a longitudinal superlattice structure. The distance between two In-O layers was about 24 Å and around 8 In/ZnO layers were interleaved between them. The spacing between an In layer and an adjacent Zn layer was around 0.31 nm, which was considerably larger than the ZnO(0002) interplanar distance of 0.26 nm. This was consistent with observations in films. Further, the authors were also able to detect nanowires with a *transversal superlattice structure*, where the layers stack perpendicular to the growth direction. The structure of these transversal superlattices was explored using HRTEM and corresponding SAED pattern, and the layers were all stacked parallel to ZnO(0002) direction. A model based on the results showed a structure similar to the longitudinal superlattice structure. The distance between adjacent In-O layers was ≈19 Å, giving an m value of 5 and the resultant formula was $In_2O_3(ZnO)_5$.

The relative ease of obtaining superlattice nanowires by self-assembly ignited the imagination of many researchers in the field, and expectations for superior devices based on these structures motivated several device studies. Several groups doped In-Zn-O nanowires with elements such as Sn, Si etc. in an attempt to improve the material properties, such as electron mobility. One such report was by Na *at al.* in 2005, who reported synthesis of Sn-doped $In_2O_3(ZnO)_4$ and $In_2O_3(ZnO)_5$ superlattice nanowires.[45] The first main observation in their synthesis comes from scanning electron microscope (SEM) images, which show a modulated nanowire diameter ranging from 50 to 90 nm. This was a deviation from the previously reported smooth-edged morphology of nanowires reported by Jie *at al.* X-ray photoelectron spectroscopy and Energy-dispersive X-ray spectroscopy measurements confirmed Sn concentration of about 8%. As hypothesized earlier, the composition of the wires varied with growth temperature. A sample grown at 900°C was reported to follow the formula $In_2O_3(ZnO)_5$ while one grown at 1000°C followed the composition $In_2O_3(ZnO)_4$. The atomically-resolved HRTEM image of a nanowire revealed the positions of In and Zn atoms directly. The results were consistent with previous reports, and the In atoms in the In-O layers were observed to occupy octahedral positions while the In/Sn and Zn atoms in the Sn-doped In/Zn-O layers randomly occupied tetrahedral and trigonal bipyramidal lattice sites (we follow the usage of the term doping by the authors, although the concentration exceeds the typical range of doping by many orders of magnitude). Polarity inversion was observed between adjacent ZnO slabs with In-O layers acting as inversion domain boundaries. The authors hypothesized that the inversion may have been facilitated by the incorporation of In atoms in the ZnO slab. Since In/Zn-O layers also contain Sn atoms, it was further suggested that Sn atoms, as well, could facilitate polarity inversion. Electron energy loss spectroscopy of $In_2O_3(ZnO)_4$ nanowire sample was carried out and elemental mapping of In was obtained from the energy loss of M shell edges (E = 443.1 eV). The results show a periodic presence of In along the entire nanowire and an alternate stacking of In-O and Zn-O slabs. This report confirmed the location of the In and Zn atoms in the lattice and also the effect of increasing synthesis temperature on the structure of the nanowire.

In 2006, Xu *at al.* reported the synthesis of In-Zn-O nanowires, where some of the nanowires showed periodical twinning structures.[46] Twin boundaries are considered as stacking faults which can alter the structure locally and mimic a heterostructure even in a homogenous nanowire due to difference in the crystal structure at the boundary. Interestingly, twinning seems to have been reported mostly in the zinc blende structure along its polar axis (111). Zinc blende ZnO is known to show twinning behavior along its (111) axis. The authors also report a zinc blende structure showing periodic twin boundaries along the (111) direction with an average periodicity of 100 nm. From their TEM analysis, it was seen that about 20-40% of nanowires show a single twin configuration with diameters ranging from 80 to 120 nm. TEM images showed a zigzag periodicity of the corrugated surface modulating the crystal diameter (here the zigzag relates to the surface shape rather than the inversion boundaries). SAED patterns from the nanowires provided additional evidence for the presence of (111) twinning structure as commonly observed in zinc blende structured crystals. From HRTEM images, it was calculated that the zigzagged surface angles were around 141, and since the (111) twinning angle in equilibrium is ≈70.53°, the zigzag structure formed along the (111) direction. The lattice spacing provided another confirmation of the same, as the distance between two adjacent planes was calculated to be 0.2673 nm, matching the (111) interplanar distance. The authors did not observe such twinning in ZnO alone and this led them to surmise that the introduction of In into the lattice is the cause for the observed twinning.

Many of the papers on In-Zn-O nanowires report studies of synthesis.[47,48,49,50,51,52,53,54,55,56,57,58,59] Some report similar observation in nanobelts,[60,61,62,63,64,65,66,67] as well as several other nanostructures.[55,68,69,70,71]

## 2.3 Other Material Systems

Attempts to diversify the material choice started already with the early ceramic thin film studies. The wide interest in superlattice nanowires attracted further research that naturally explored other materials in the hope to increase the material choice and engineer certain material properties. Since the In-Zn-O system has been the easiest venue to obtain the superlattice, many of the studies only slightly perturbed that system by introducing "doping" of additional metals. This so-called "doping" was typically at the percent range rather than the part-per-billion that is typical of semiconductor doping. In many of these cases, the zigzag boundary is still observed. However, in some of the cases the boundaries are typically flat while the periodic polarity inversion is common to all of these systems.

### (1) Adding a metal to the In-Zn-O system

The smallest perturbation to the In-Zn-O system was perhaps the addition of another group III metal. Growth of In-Ga-Zn-O nanowires was first attempted by previously-mentioned Nakamura *at al.* and Moriga *at al.* already in ceramic thin films,[32,33] and used by Nomura *at al.* to make thin film transistors.[13,14,15] More recent work on ceramic thin film was carried out by Kim *at al.*[72], Narendranath *at al.*[73], and Huang *at al.*[74], the latter reporting reduced lattice thermal conductivity induced by the introduction of Ga. In-Ga-Zn-O nanowire growth was reported by Li *at al.*[75], Guo *at al.*[76], Lou *at al.*[77], Clairvaux Felizco *at al.*[78] The same structure was also reported in nanobelts by Wu *at al.*[79]

Most of the work on In-Al-Zn-O was carried out on ceramic thin films, starting with the work of Nakamura *at al.*[80] who reported the phase relations. Other more recent works on films explored the improvement in thermoelectric properties.[81,82,83] In-Al-Zn-O nanowire synthesis was reported by Huang *at al.*[84]

Introduction of Fe into the In-Zn-O system was studied only in ceramic thin films. It was first attempted by Kimizuka *at al.*[85,86,87,88] and by Uchida *at al.* (same group).[89] Later Hörlin *at al.* studied the zigzag patterns defining them as defects.[90] More recent works by van Erichsen *at al.*[91] studied synthesis and structure, and a work by Zhang *at al.* studied thermal transport in films of this composition.[92]

Park *at al.* studied the effect of an added Sb on the electrical conductivity and crystalline structure of the In-Zn-O. [93]

Sn was originally introduced by Palmer *at al.* into In-Zn-O ceramic layers to improve the properties as a transparent conductive layer.[94] More recent works on ceramics from the same group studied the phase relationship and the synthesis.[95,96] These were followed by studies of nanowire growth.[97,98]

Finally, addition of Yt was reported in two studies of ceramics to improve thermoelectric properties.[99,100]

### (2) Replacing In by another metal

Ga-Zn-O was first proposed as a new homologous series by Kimizuka *at al.* in 1995,[38] and Li *at al.* in 1999.[101] While Ga is a group III metal as In, HRTEM study by Barf *at al.* clearly shows that both inversion domain boundaries of each ZnO section are flat (no "zigzag" boundary).[102] The $Ga_2O_3(ZnO)_m$ homologous superlattice compounds are a distinct type of structure series different from other ZnO-based superlattices such as $In_2O_3(ZnO)_m$. $Ga_2O_3(ZnO)_m$ crystallizes into an orthorhombic structure with a Cmcm space group. The $Ga_2O_3(ZnO)_m$ structure can be developed by stacking the consecutive slab-like structure units, which are related to each other by mirror symmetry.[103] The mirror boundary serves as a twinning operation, leading to a kind of unit-cell twinning. The $Ga2O3(ZnO)m$ structure consists of Ga-O and (m+1) Zn/Ga-O layers stacked alternatively along the c-axis.[104] $Ga^{3+}$ ions are located midway between the two adjacent mirror boundaries with fivefold bipyramidal or trigonal coordination, which apparently differs from those in $In_2O_3(ZnO)m$ in which In atoms occupy the octahedral sites. Across the Ga-O layer, the $ZnO_4$ tetrahedra are inverted. Electrical and thermoelectric properties of Ga-Zn-O films were reported by Michiue *at al.*, Yoon *at al.*, and Guilmeau *at al.*[105,106,107] Transport and optical properties were reported in Ga-Zn-O nanowires by Yuan *at al.* and Chang *at al.*[108,109]

To our knowledge, no $Al_2O_3(ZnO)_m$ homologous superlattices with evidence from transmission electron microscopy have been reported so far. Fe-Zn-O study were first reported by Li *at al.* following a work on a Lu-Fe-Zn-O by the same group.[110,111] They showed that the Fe-Zn-O system forms zigzag domain inversion boundaries like the In-Zn-O system. The same zigzag boundary structure was also reported in two works by the same group, Köster-Scherger *at al.*[112] and Wolf *at al.*,[113] while microwave synthesis was reported by Nagao *at al.*[114] $Fe_2O_3(ZnO)_m$ form a superlattice structure that is isostructural with $In_2O_3(ZnO)m$. However, the $Fe_2O_3(ZnO)_m$ superlattices usually develop at low iron concentration or for high m values like 15.[115,116,117] Superlattice structures of lower order, such as with m = 4, 7, 9, have not yet been experimentally observed. For high Fe concentrations, $ZnFe_2O_4$ spinel precipitates in morphologies of nanolaminate and nano-spheres.[118]

Cd-Zn-O superlattices were synthesized and photoluminescence was studied in nanowires by Lopez-Ponce *at al.*[119]

Sb-Zn-O was shown by Park *at al.* to possess periodic twinning with flat twin boundaries.[93]

Finally, Cao *at al.* observed the superlattice structure in Sn-Zn-O nanowires. They also observed a significant reduction in electrical conductivity when measured perpendicular to the periodic boundaries as compared to measurements parallel to the boundaries.[120] Periodically twinned Sn-Zn-O nanowires were also reported by Kim *at al.*[121]

### (3) Binary Systems

As previously discussed, these homologous self-assembled superlattice structures have been found to be consisted of periodic sequences of polarity-inverted domains, where each section is, in principle, identical to its two neighboring sections, except for its inverted polarity. Several reports suggest that periodic polarity inversion does not necessarily require the presence of another metal, as it has been also observed in undoped ZnO,[122] as well as in other undoped binary wurtzite crystals, e.g., ZnS,[123] and SiC,[124,125] where it is commonly referred to as periodic twinning.

## 2.4 Theoretical Studies

Theoretical studies of these superlattices mainly address two general aspects: Crystallographic structure, and electrical/optical properties. All these works take as granted that the ZnO flips its polarity in a periodic manner and that the other metal is found at the IDBs. The structural studies mainly address the formation of the zigzag boundary and convincingly model its formation using structural energy considerations.

### 2.4.1 Zigzag Inversion Domain Boundary

While the zigzagged boundary was common in most of the early reports, it was generally missing in most of the more recent reports. In an attempt to shed more light on this apparent inconsistency, Yan *at al.* performed DFT calculations of the atomic structure of the lattice.[126] The focus of their study is the nature of the inversion domain boundaries since structural characterizations have indicated towards the presence of a second layer within the ZnO slab which acts as a mirror domain boundary. Based on total energy calculations, they find that a zigzag modulated boundary inside the (Zn/In)O slabs which should have a lower energy than a flat layer by as much as 4.24 eV. To further explain why the flat layer is not energetically favorable, the authors offer two explanations; the first being that the lattice mismatch between a flat In-O layer and a ZnO atomic layer is too high, causing considerable strain around the In-O boundary. Secondly, O atoms are threefold coordinated in the flat layer arrangement which is not energetically suitable. These two constraints are absent in the zigzag form, wherein the O atoms are all fourfold coordinated and the atomic order is alternative, causing little strain around the boundary. To explain the observations of flat boundary layers, they suggest that these are more likely to form in cases of short ZnO sections, while the zigzag is more likely in longer ones. The authors offer an explanation based on the assumption that thin regions will always show a flat layer while images taken on a thicker layer will reveal the modulated structure. Using Simulations of the HRTEM images, they show that the In-O octahedral layer exhibits clear contrast in the image due to absence of Zn atoms and a different coordination (sixfold) compared to that of the ZnO slab (fourfold). In comparison, the modulated In-O boundary consists of a mix of In and Zn atoms and the local atomic arrangement is close to that of ZnO, leading to no clear contrast. Hence, in real samples, because of increase in the number of kinks, one can see the modulated structure in thicker samples, but not in thinner ones. Electron diffraction patterns simulated on the two arrangements showed extra spots when the zigzag modulation is considered. Such a pattern has been experimentally verified.

The structure of homologous series with the formula $R_2O_3(ZnO)_3$ (R = Al, Ga, and In) was studied around the same time by Yoshioka *at al.* [127] They, however, based their study on models describing the coordination numbers in different atomic arrangements and compared their results with those of Schinzer *at al.*[39] They found that although the $R_2O_3(ZnO)_3$ phases all show positive formation energies, the contradiction can be resolved using finite temperature effects. In their model, there is a strong preference for a fivefold coordination in $Al_2O_3(ZnO)_3$ and $Ga_2O_3(ZnO)_3$, but the formation energies are independent of the choice of In or Zn atoms, which is in line with the fact that In and Zn occupy same sites and are randomly distributed.

An extensive study proposing rules for structure formation of In$MO_3$(ZnO)$_n$ compounds ($M$ = In, Ga, Al; $n$ = integer) was reported by Da silva *at al.* [128] in 2008. Starting with the results of Yan *at al.*, the authors pose the question as to what exactly is the ground state configuration of such compounds and the underlying mechanisms of its stability. Using first-principle calculations they propose several configuration rules according to which the crystal energy is lowered. These rules include the octahedron rule for the InO$_2$ layers, presence of IDBs maximized hexagonality in the ($M$Zn$_n$)O$_{n+1}$ layers [64], minimum strain in the interface between the InO$_2$ and ($M$Zn$_n$)O$_{n+1}$ layers and the electron octet rule.

The authors delineate rules for obtaining the configuration with the smallest energy, and present models of the ground state configuration of In$MO_3$(ZnO)$_n$ including the modulated structure that follows all the rules; an inversion domain boundary, tendency to preserve hexagonality in the ($M$Zn$_n$)O$_{n+1}$ layers and arrangement of Zn and $M$ as trigonal bipyramids, obedience of the octet rule and the formation of the modulated zigzag structure. They show that their ground state configurations and rules may be applied to a wide variety of these homologous structures providing a comprehensive model for the observed structures of the inversion domain boundaries.

In another work from the same group, Da Silva *at al.* [129] used DFT to study the $Ga_2O_3(ZnO)_6$ system. Their results explain the formation of zigzag boundaries in this system. They also study the electrical properties and find them to be determined mainly by the electronic properties of ZnO with minor changes in bandgap and optical absorption.

Michiue and Kimizuka [130] propose a unified description for $Ga_2O_3(ZnO)_m$ structures using the superspace formalism which accounts for the zigzag modulated boundaries.

In contrast, Röder *at al.* [131] propose a model based on DFT of the $In_2O_3(ZnO)_7$ system without considering the zigzag boundary but nontheless find it to match well with HRTEM images.

The results of Yan *at al.* were shown to have a few drawbacks by Wen *at al.* [132] who point out that the modulated structure can only be distinguished in a single direction which depends on the arrangements of M and Zn atoms in a given projected direction, which cannot eliminate the uncertainty in the positions of the M atoms. Further, they suggest based on experimental evidence that the zigzag pattern is observed when the observed directions are oriented to the equivalent [010] and [$\bar{1}$10] directions of the hexagonal lattice which also cannot be explained by the model suggested by Yan *at al.*

In their report, Wen *at al.* show a 2*m* x 2*m* supercell with a specific atomic arrangement that should be taken as the IMZO unit cell. HRTEM simulations were carried out using a multislice method choosing $In_2O_3(ZnO)_3$ and $In_2O_3(ZnO)_6$ as representative and generalized cases. The authors propose a crystal structure having a zigzag boundary formed due to a unique arrangement of In atoms within the In/Zn-O slab projected along the ($\bar{1}$10) direction of the hexagonal lattice. The unit cell is said to be comprised of two such structures stacked along the c axis. A total of 288 In atoms are present in this structure with half of them occupying the octahedral sites constituting the $InO_2^-$ layer, while 33 are at the trigonal bipyramidal and tetragonal sites of the In/Zn-O slab, forming the zigzag boundary. The structure does not require a fivefold coordination for the In atoms and the periodicity of the zigzag shape may be obtained from the relation $T_{[-110]} = ma$ along (110) direction and $T_{[010]} = \sqrt{3}ma$ along the (210) direction. This relation explains the observation of linear relationship of the zigzag periodicity with *m* as observed by Li *at al.*[10,41]. The authors also suggest a formula for calculating the apex angles for the zigzag shapes projected along [$\bar{1}$10] as $\alpha = 2\arctan[\beta(m+2)a/c]$ where $\beta = 1$ when *m* is even and 3/2 when *m* is odd. Thickness of the In/Zn-O slab is c/2 for even *m* and c/3 for odd; hence the parameter $\beta$ is added to distinguish the difference. For the other direction [010], the parameter changes to $\sqrt{3}\beta$. The difference between the calculated and measured angles are hypothesized to arise from experimental measurement errors and lattice distortions. Simulated HRTEM images are then obtained using the model described above for the unit cell corresponding to $In_2O_3(ZnO)_6$. Interestingly, only the upper half of the figure contained the pattern. Experimental results have previously reported that for values of *m* smaller than 6, the zigzag pattern is not clearly identified.[10,41] For large values of m, the image contrast at thicker regions is enhanced when viewed along ($\bar{1}$10) [10]. It was also observed that when the image is taken along (010), the zigzag pattern disappears in the regions with variable thickness. Invisibility of the modulated structure with small *m* is explained by a small periodicity of the zigzag shape, causing the diffracted electron beams to interfere with each other, resulting in indiscernible zigzag contrast of the image. From the schematic diagram suggested earlier, the authors interpret the direction dependent visibility by the number of atoms along that direction. When the image is observed from a direction parallel to ($\bar{1}$10), it is enhanced since the zigzag boundary is primarily due to the In atoms and hence can be intensified by their larger numbers along this direction. Similarly, along the (010) direction, it can be deduced that the most clear image would be obtained if the thickness of the specimen has a value of (2k+1)*ma*. It becomes difficult to distinguish the image if the thickness is 2*kma*. Simulating the HRTEM image of $In_2O_3(ZnO)_6$ with the incident beam along [010] with two different thicknesses in the specimen shows that the upper part exhibits a clearer zigzag pattern compared to the lower one. Therefore, it is due to variable thickness that some experimental images of the $In_2O_3(ZnO)_6$ do not show the zigzag pattern [85] and the visibility of the zigzag pattern varies with thickness of the sample in the case of $InFeO_3(ZnO)_{13}$ compound [41]. Next, the authors calculate the formation energies of the $InMO_3(ZnO)$ compounds comparing different structural models and they show that their proposed model possess the lowest total energy and is thus claimed to be the correct ground state configuration of these materials.

In a more recent paper from the same authors, Wen *at al.* [133] compare several models for the $In_2O_3(ZnO)_m$ system and find that the zigzag boundary produces the greatest stability, and that the bandgaps and effective masses increase with *m*.

Considerable theoretical effort as so far been made to understand the zigzagged modulated inversion boundary. While this peculiar boundary structure impeded progress in understanding of this self-assembled superlattices for many years, it is by no means a common denominator to all of these systems.

### 2.4.2 Electrical Properties

Several theoretical efforts dealt with the anisotropic electrical conductivity, the dependence of the electrical conductivity on structural parameters, the defects responsible for this conductivity, and the optical band-to-band transitions (sometimes referred to as "bandgaps"). The small number of these studies stands in contrast to their crucial importance for utilizing these structures in practical applications.

Walsh *at al.* [134] used DFT to calculate band structure and attempted to explain the electronic properties of $In_2O_3(ZnO)_n$ compounds (*n* = 1, 3, 5) by analyzing Zn rich and Zn poor versions of the compound. They proposed explanations for the redshift observed for ternary compounds in terms of forbidden transitions being overcome due to the formation of a superlattice. Further, it explains why Zn poor compounds display optimal conductivity, as localization on $InO_2$ octahedra increases with the Zn concentration.

Yoshinari *at al.* [135] studied the crystal structures of several of these homologous compounds, namely $Zn_3In_2O_6$, $Zn_4In_2O_7$, $Zn_5In_2O_8$, $Zn_7In_2O_{10}$ and $In_2O_3$ by X-ray Rietveld analysis. First principle calculations were carried out to evaluate the band structures using data obtained from the refinement. For *k* = 3, 5 and 7, a single phase of the *R3-m* type $Zn_kIn_2O_{k+3}$ was observed while for *k* = 4, it was the *P6₃/mmc* type. The basic structure was reaffirmed to be $InO_2^-$ layers alternately stacked with $(InZn_k)O_{k+1}^-$ with edge sharing $InO_6$ octahedra forming a single layer on the c-plane. In the vicinity of the center of the $(ZnIn)_kO_{k+1}^+$ layers, the In or Zn atom gets displaced from the central site of tetrahedron in a direction opposite to the apical oxygen and forms a trigonal bipyramid with the basal oxygen atoms. Plotting In-O interatomic distances and O-In-O angles for $Zn_kIn_2O_{k+3}$ shows that with decreasing values of *k*, the In-O distance increases and the angle

O-In-O gets closer to 90°, pertaining to a regular octahedron. $Zn_4In_2O_7$ is an exception probably due to difference in space groups. Comparing calculated electronic structures, they were also able to explain an observed difference in electronic conduction between $Zn_3In_2O_6(4)$ and $Zn_3In_2O_6(5)$ in a difference in the calculated conduction in the In-In chain.

Peng *at al.* have used the structure models detailed by Da Silva *at al.* to study defect formation energies in order to find the carrier generation mechanism.[136] It was observed in experiments that $In_2O_3(ZnO)_k$ showed anisotropic electrical conductivity.[137] However, theoretical calculations using LDA approximations show similar dispersion in the lowest conduction band for k = 1, 2 and 3 in the three directions along the reciprocal lattice vectors. In this study, the anisotropy is then suggested to arise from spatial distribution of defects. Weighing in the possible contribution of various defects the authors suggest that the In-O layer should be the primary in-plane conducting path while the zigzag boundaries the corresponding out of plane path. At equilibrium, $V_O$ occurs along the zigzag path while the $In_{Zn}$ also forms complexes with $V_O$ and is found on the same zigzag boundary. The main out-of-plane conducting path is then blocked by the high concentration of scattering centers with high deformation potentials ($V_O$) or strong Coulomb potentials ($In_{Zn}$) [138] because of low values of the dielectric constant. These phenomena are suggested to lead, eventually, to the strong anisotropic electrical conductivities in $In_2O_3(ZnO)_k$.

As we will argue later, the band structure perpendicular to the inversion domain boundaries appears to oppose electron conduction altogether. Thus, the material in this direction should behave as an insulator, which suggests that measurements of currents in that direction where probably reflecting grain surface currents.

## 2.5 Applications

Several possible applications have motivated the research of these homologous compound structured material. As we are interested in the structure, we will only mention the applications briefly and refer the interested reader to published reviews.

### 2.5.1 Varistors

Some of the earliest works on these structures were on Sb-doped ZnO that is commonly used for varistors. The first report on "doped" ZnO for varistors dates back to 1969.[139] The early work on materials for varistors has been thoroughly reviewed by Clarke.[140] Varistors are typically based on ZnO ceramics that includes additional oxide or oxides. Most of these works are in the spinel range, while those in the zincite range are typically not concerned with the fine structure, and the superlattice structure was reported to form only with few of the studied varistor materials.

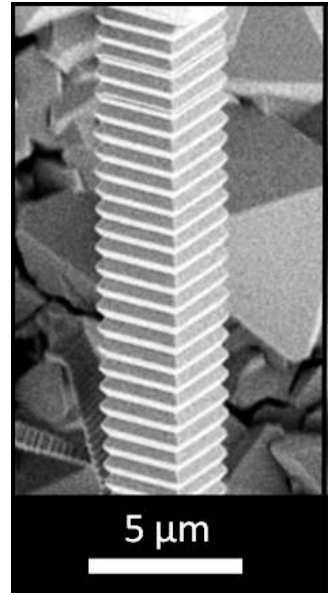

**Fig. 4** Periodically twinned micro-rod that we observed unexpectedly in an attempt to grow $In_2O_3$ nanowires. In thick wires and rods, the surface is periodically corrugated making it easy to identify the periodic twinning.

### 2.5.2 Transistors

The performance of a field-effect transistor has become a popular means to evaluate a semiconductor material or structure for electronic applications. We have already mentioned in the introduction the works on thin-film transistors that followed the work of Nomura *at al.*[13-21] some which reporting relatively high channel mobilities (greater than generally reported in single crystal ZnO).[16,20] Nonetheless, these materials have not made it into the realm of electronics. In polycrystalline materials, grain boundary scattering generally reduce the mobilities, and therefore they hardly ever become a material of choice for a transistor channel. In contrast, nanowires are single crystals, and should be expected to show greater channel mobilities. Hsu and Tsai indeed achieved 86 $cm^2V^{-1}s^{-1}$ in In-Zn-O nanowire channel field effect transistors.[141] This results is still quite remarkable considering the fact that even in the single crystal form, the inversion domain boundaries are known to be very efficient electron scaterers, suggesting that the transport was probably along the boundaries within the boundary quantum wells.

Notwithstanding, these mobilities are no match to the state-of-the-art channel mobility in electronic devices nowadays which is at a couple of thousands of $cm^2V^{-1}s^{-1}$ obtained in III-V quantum-wells high electron mobility transistors. Rather, they serve as a certain figure of merit for the electrical conductivity of these material structures.

### 2.5.3 Thermoelectrics

In contrast to the good electrical conductivity, the phonon conductivity across the inversion boundaries is relatively low.[142] The combination of good electrical conductivity (electrons) and poor heat conductivity (phonons) is a desired combination for thermoelectric conversion.[143] This combination places the greatest promise of these homologous self-assembled superlattice compounds in thermoelectric applications. This potential was first recognized in papers by Koumoto group, Nagoya University,[144,145,146,147,148,149,150,151] evaluating the thermoelectric properties in $In_2O_3(ZnO)_k$ of various k values. In the past decade, publications from several groups marked a renewed interest in these materials for thermionic applications, of which noteworthy is a set of rigorous studies by Liang *at al.*[152,153,154,155,156,157] The interested reader is referred to a recent comprehensive review of the subject by Liang.[158]

## III. MODEL

In this section, we build upon the foundations laid by the accumulated knowledge that we have reviewed so far to propose a mechanism to explain the observed self-assembly of the homologous superlattices. To enable electrostatic effects on crystal growth, the crystal has to be able to interact with electric charges or electric fields. Such interaction has been reported by some of the authors of this paper [159,160] and by others [161] for wurtzite materials. Wurtzite crystals have a built-in electric field that can interact with mobile electric charges. This field is a natural trait of the wurtzite structure, a result of its spontaneous polarization. For example, ZnO nanowires often grow vertically-aligned on $SiO_2$ substrates. $SiO_2$ is amorphous, and thus cannot provide epitaxial guidance to align the wires. Nonetheless, the ZnO wurtzite lattice has an intrinsic built-in electric field, and it was shown that the interaction of this field with the field emanating from parasitic charge in the $SiO_2$ causes the alignment of the polar axis of the crystal with the external field. When the sign of the electric charge in the $SiO_2$ was changed, the polar axis of the wires flipped over.[159] These observations showed an electrostatic effect of substrate charge. However, the same mechanism could be in effect at each inversion domain boundary along a periodically polarity-inverted wire. All that is required is that during the growth of each section, mobile electric charge will gradually accumulate at the growth edge until, at a very specific length, it will build up a strong enough electric field to cancel the effect of the polar charge at that face and thereby to enable the flip over of the polarity of the next section. The same process would then reiterate in the next section and so on and so forth. The crystal's polar built in electric field separates mobile charges along the growth axis in a way that charges of one sign float to the growth edge, while charges of the opposite sign are swept downward (Fig. 5).

Let us denote the In in the In-Zn-O as the *secondary cation*. The secondary cations observed in the homologous superlattices usually do not form oxides that have spontaneous polarization. It seems that for this reason, they have higher likelihood to attach to the growth surface, once the polar charge is fully screened out by mobile charge. When they do, their bond angles allow them to bridge over efficiently to the inverted phase that starts to grow immediately on top of them.

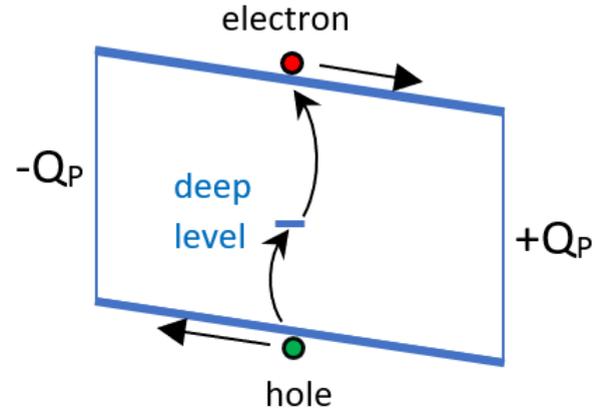

FIG 5. Band diagram of a single period of the periodic alternating-polarity structure. A single period of, e.g., ZnO, has opposite polar charges on its ends that give rise to a built-in field. The typically high temperature of the growth increase the probability of thermal generation, especially if mid-gap states such as deep levels and surface states are available and reduce the minimal energy required for band-to-band generation. The generated electron-hole pair is then separated by the built-in field.

Some of the inversion boundaries are flat (when an O-face is facing an O-face) and some of them are zigzagged (Zn-face to Zn-face boundaries are zigzagged in some of the cases). A similar zigzag inversion boundary has been reported in InN and ZnO nanorods.[160] In those systems, there were no secondary cations, and the transition took place over much longer sections of the order of more than a micron. The mechanism proposed there was a symmetric non-uniform distribution of mobile charge causing inversion only at the highly charged points in the structure. At this time, we cannot explain why there would be such a fundamental difference in the mobile charge distribution between the two polar faces of the homologous superlattice system. However, the mechanism of mobile-charge-induced polarity-inversion is definitely in line with the present model.

There could be two major sources for electron-hole thermal generation: in the bulk, and/or on the surface. Thermal generation seems likely during crystal growth as the growth temperature is typically high. However, direct band-to-band excitation is less likely in a wide gap material as is ZnO then in e,g., GaAs. To account for generation in wide gap materials such as ZnO, we have to consider an indirect excitation through deep-levels (Fig. 5). Deep levels at surfaces (surface states) typically span wide energy ranges (e.g., the yellow luminescence band in GaN,[162] or the green luminescence band in ZnO[163]) making them the more probable route for generation.

The amount of charge generated in each wire section should be relative to its volume, if the charge is generated in the bulk of the material, and should be relative to its surface area if the surface is the source. Whether it is this way or the other, the rate, at which the charge will be accumulated, has to increase with the wire radius, because both the surface area and the volume increase with the wire radius. If such charging mechanism is indeed at work, this would predict a dependence of the section length on the wire diameter. The exact nature of this dependence, however, should vary with the relative part of each of the generation mechanisms, surface or bulk.

In the following section, we show that experimental data from the literature may be used to test this explanation, and if validated, to also attest the origin (bulk vs. surface) of the generation mechanism.

## IV. PROPOSED EXPERIMENTAL TEST

Understanding of the structure of the ZnO-based homologous self-assembled superlattices has evolved along the years. Nonetheless, no clear idea of what causes the layers to self-assemble the way they do has been put forward so far. While the effect was observed in both polycrystalline films and in nanowires, we choose, for convenience, to start with modeling their formation in nanowires. The reason for this choice will be explained later, when we move on to generalize to include other structures and layers. Currently, we understand that the structure of these self-assembled superlattice nanowires is in fact very similar to that observed in periodically-twinned ZnO nanowires. The ZnO wire is periodically sectioned into identical sections having alternating polarity along the polar-axis of the crystal. Each two consecutive sections are separated by a domain-inversion boundary. In the superlattice version of this structure, an additional metal, or a combination of several metals, forms a monolayer that bridges and connects the two neighboring domains of opposite polarities right at the domain-inversion boundary. Also, in the homologous superlattice, the domain inversion boundary is not always a twinning boundary.

The consecutive sections are typically short (1 to 100 nm). Since the polar axis is always perpendicular to the inversion boundaries, each of the two ends of each single section has a constant polar charge of opposite sign. These two opposite sheet charges constitute between them a built-in electric field. Due to the short length, this field is typically very strong, tilting the energy bands (Fig. **6a**) and leaving this section fully depleted. In each section, the field is opposite to the fields in adjacent sections. Hence, a periodically-twinned ZnO nanowire may be viewed as a naturally-formed doping superlattice (Fig. **6b**), also known as a "n-i-p-i" superlattice structure,[164] where the domain or section boundaries are charged with charges of identical magnitudes and periodically alternating signs. To overcome this endless series of barriers and induce conduction across them would require a very high voltage if at all possible. This suggest that the transport measurements cited at the end of section 2.4.2 actually measured conduction at grain boundaries rather than within the grains.

When another metal, e.g., indium, is added to the structure, no ternary phase is formed as long as the In density is small, but rather the other metal forms a single atomic layer (not even a monolayer) that bridges between the two adjacent opposite polarity slabs of ZnO. Perhaps this other metal may be viewed as a "polarity-inversion-promoter", although periodic-twinning appears to take place also in the absence of the other metal. The presence of this other metal oxide at the domain inversion boundary probably perturbs the energy band line up creating notches, quantum wells and quantum barriers, at the very boundary (Fig. **6c**), because of the band discontinuity between the ZnO and the oxide formed by the other metal.

We have previously reported the effect of charging and electric fields on the growth direction of polar semiconductors.[159,160] In the case of polar semiconductors, especially ZnO, it appears that the effect of electric fields emanating from charges may sometimes have a stronger effect than epitaxial guidance.[159] If a nanowire of polar semiconductor is growing in the polar direction, fixed polar charge will always be present at the growth end of the wire to guide further growth. However, this fixed polar charge is often partially or totally screened out by mobile charges of the opposite sign.

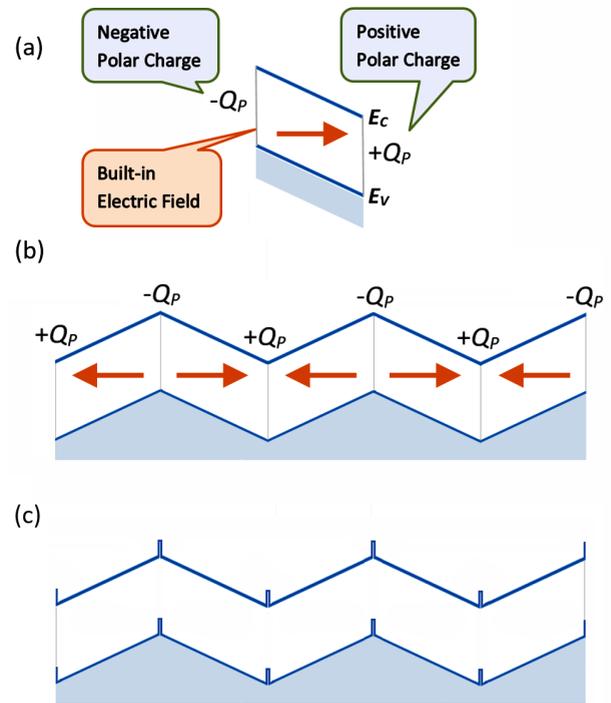

**FIG 6**. Qualitative electronic band diagrams of an In-Zn-O homologous superlattice. (a) A single ZnO section – the intrinsic polarity induces a built-in electric field that tilts the bands. (b) Periodically twinned ZnO – essentially possesses a band structure that is similar to n-i-p-i superlattice consisting of an alternating series of triangular potential wells and triangular barriers. The built-in electric field in each section as an opposite direction than the fields in the two adjacent sections. (c) In-Zn-O – if the added indium were $In_2O_3$ it would probably create an off set quantum size layer as shown. In reality the layer is $InO_2$, but according to photoluminesnce studies, the offset is more or less the same.

Thakur *at al.*

Let us imagine a nanowire growing along the polar axis. The growth takes place at the upper edge of the wire forming a new section/domain. The growth is carried out at a high temperature. Energetic phonons give rise to thermal generation of electron-hole pairs in this upper section as it is formed. The strong built-in polar electric field separates the pairs. Electrons are swept to the end with the positive polar charge, while holes are swept to the end with the negative polar charge. After their life time is over, the charges recombine. The balance between generation and recombination leaves a certain net charge that is swept to the growth edge and cancels some of the fixed polar charge at that end. We will assume that this band-to-band generation takes place in the bulk of the material, and its rate is fixed per unit volume, while in parallel, there is also an indirect generation through bulk deep levels at a fixed rate per unit volume, and an indirect generation through surface states takes place on the side walls of the growing section and its rate is fixed per unit surface area.

As the section grows longer, its volume increases and so does its surface area, though not at the same rate. This brings a gradual growth in the net generation. The growth continues until a critical length is reached, at which the generated mobile charge accumulated at the growth end totally screens the polar charge. Passing this point induces a flip-over of the growth polarity and an inversion boundary is formed. A new section starts to grow with an opposite polarity, and again, its growth is limited to the same critical length for the very same reason. The flip-over is assisted by atoms of the other metal that can more easily form the slightly different angles between the chemical bonds that are required for the polarity flip-over, but the flip-over may also take place without the other metal, thus forming a periodically domain-inverted *homostructure* wire rather than the *heterostructure* superlattice wire.

The above mechanism implies dependence of the section length (the period of the superlattice) on the wire diameter as well. On the other hand, it does not imply any dependence on the type of the other metal or its presence at all.

If we assume a net density of pairs generated (whether direct or indirect) in the bulk (generation minus recombination) of $G_B$, then to get the number of generated pairs we will need to multiply it by the section volume, $\pi R^2 L$, where $R$ is the radius of the wire, and $L$ – the section or period length. Similarly, we will need to assume a net generation at the surface, $G_S$, and multiply it by the section surface area, $2\pi R L$, to get the number. Adding these two numbers, we get the critical number of charges required to induce a polarity flip-over. This critical number equals a density, $C$, times the wire cross-section area, $\pi R^2$.

$$\pi R^2 L G_B + 2\pi R L G_S = \pi R^2 C \qquad (1)$$

Extracting the section length, $L$, we get a dependence of the section length on the nanowire radius, $R$:

$$L(R) = \frac{C}{G_B}\frac{R}{R + 2\frac{G_S}{G_B}} \qquad (2)$$

To test the validity of this model, we need data points of (R,L), where each superlattice wire provides a single data point. Fitting the data with Eq. 2 should tell us about the ratios among the 3 parameters. Direct generation in the bulk requires that a phonon (~0.1 eV on the average for the typical growth temperature) will excite an electron over the bandgap. In the case of ZnO the bandgap is 3.3 eV. The probability for this transition should be extremely low. In contrast, deep levels, whether on the surface or in the bulk situated somewhere around the center of the energy gap may reduce the required energy leap by as much as half, making such generation more probable by several orders of magnitude. Since surface states typically form extremely wide bands that may extend over more then 1 eV, the probability of the surface excitation appears more likely. Figure **8** shows for example a photoluminescence spectrum of In-Zn-O homologous superlattice nanowires. While the band-edge is lower by 0.1 eV relative to its typical position in ZnO, the sub-bandgap spectrum shows the classic sub-bandgap PL of ZnO: the so-called green luminescence band that spreads over ~1.25 eV of the bandgap energy. The ZnO green luminescence has been positively identified to be a surface state (See, for example, Ref. and references therein). Hence, it is reasonable to expect that $G_B \ll G_S$. In such case, one may omit the bulk generation term from Eq. 1, and obtain the following linear dependence of the section length on the nanowire diameter:

$$L(R) = \frac{C}{2G_S}R \qquad (3)$$

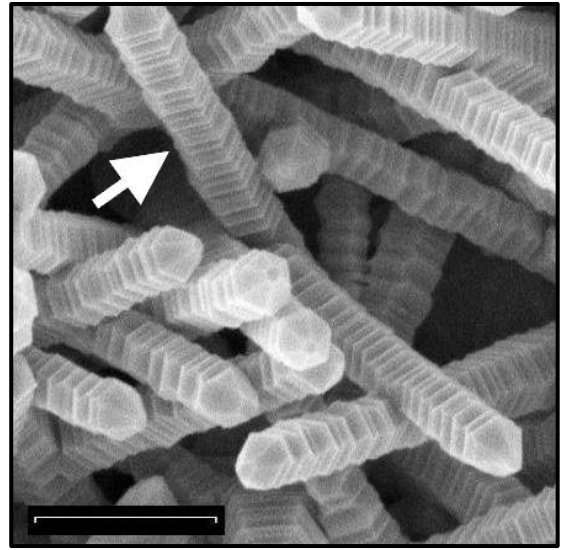

**FIG 7**. Data point 9 in table I was obtained from this SEM image from the wire pointed at by the arrow. The scale bar is 1 micron long.

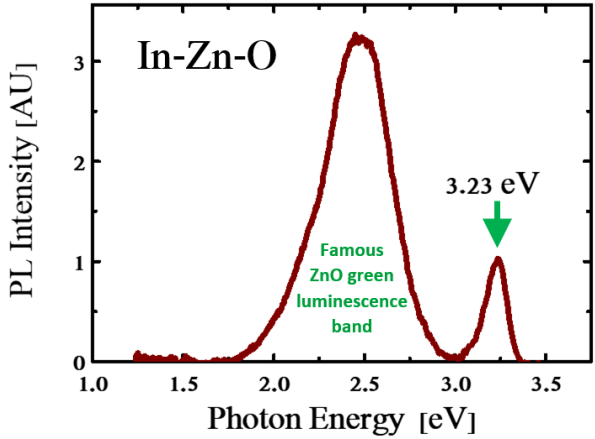

**FIG 8**. Photoluminescence obtained from In-Zn-O homologous superlattice nanowires. Compared with ZnO, the band-edge is red-shifted by 0.1 eV. However at sub-bandgap energies, the famous ZnO green luminescence band, known to be a surface state [See Ref. 122 and references therein], spreads over a range as wide as 1.25 eV. This is not surprizing as the main surface area of the wire is that of the ZnO sections.

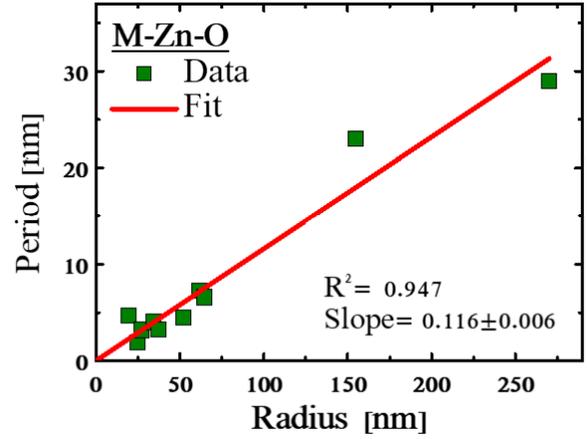

**FIG 9**. Period length as a function of wire radius for M-Zn-O superlattice wires which data is given in Table I (green squares) and a fitting curve obtained using using Eq. 3 (red line). It should be emphasized that the data were collected from the literature from various studies and growth carried out under different conditions, but variance is nonetheless excellent.

TABLE I. Period (section length) and radius data of superlattice M-Zn-O nanowires collected from the literature.

| Data | M | L[nm] | R[nm] | L/R | Reference |
|---|---|---|---|---|---|
| 1 | In, Sn | 1.9 | 25 | 0.076 | Na 2005 |
| 2 | In, Al | 3.2 | 27.3 | 0.117 | Huang 2010 |
| 3 | In | 3.3 | 37.5 | 0.088 | Niu 2010 |
| 4 | Sn | 4.0 | 34.3 | 0.117 | Cao 2012 |
| 5 | In | 4.5 | 52 | 0.087 | Jie 2004 |
| 6 | In, Al | 4.7 | 19.3 | 0.244 | Huang 2010 |
| 7 | In, Al | 6.6 | 64.5 | 0.102 | Huang 2010 |
| 8 | In | 7.3 | 61.5 | 0.119 | Zhang 2008 |
| 9 | In | 23 | 155 | 0.143 | This work |
| 10 | None | 29 | 270 | 0.107 | Shalish 2004 |

L-R Correlation = 0.97

## V. MATERIALS AND METHODS

To test the model, we used superlattice nanowire data that we collected from the literature. This was because we found our own data to span only a small part of the radius range reported in the literature. Our preference was not to use numbers provided by authors but rather to assess all the data from the published electron microscopy images. To be included in our statistics, both the length of the section/period and the radius of the wire had to be unambiguously definable from electron microscope images.

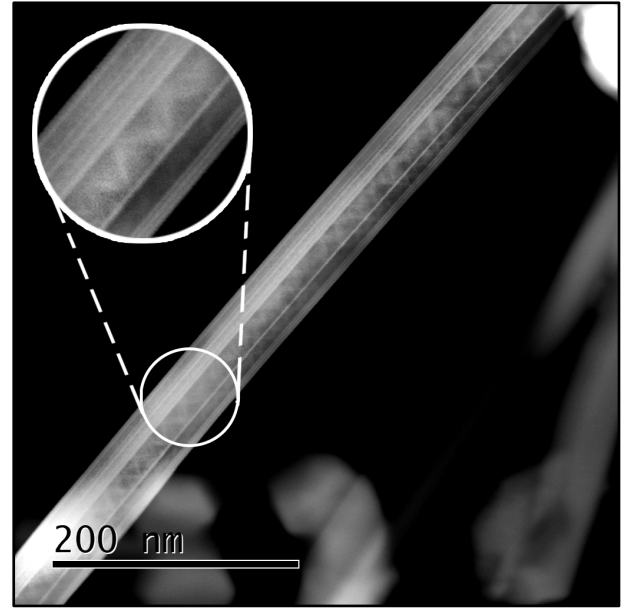

**FIG 10**. In-Zn-O wire wherein the periodicity is across the wire. The inset magnifies a portion of the wire showing the zigzag domain inversion boundary as commonly observed in the In-Zn-O structures.

The 9th data point for ZnO periodic lattice (without additional metal) was obtained from nanowires (Fig. 7) grown by chemical vapor deposition on Si(111) wafers from a source (mixture of ZnO and graphite powders with indium metal) held at 1030 °C, while the substrate was held downstream at a point where the temperature was 950 °C. The growth was carried out under a flow of 30 sccm of Ar and lasted 30 min.

## VI. RESULTS AND DISCUSSION

We were able to find 10 such data points of M-Zn-O superlattice nanowires that met our selection criteria. The data and sources are listed in table I. SEM image of the wires used for data point 9 is shown in Fig. 7. The 10$^{th}$ data point in Table I is a case of periodic twinning of ZnO. In that case, the same periodic structure was formed without intentional addition of another metal. Correlation of 0.97 calculated between the period, L, and the radius, R, for these data, corroborates our hypothesis of causative connection between the two parameters. Fit of these data with Eq. 2 yielded a ratio $G_S/G_B \cong 10^{14}$ as predicted, meaning that the net generation in the bulk was extremely negligible (practically zero) compared to generation at the surface. We, therefore, replaced Eq. 2 with Eq. 3.

Figure 9 plots the period, L, as a function of the radius, R, for the data of Table I, and the fitting curve calculated using Eq. 3. The calculated coefficient of determination, $R^2$, for the fit is *0.95*. This high value suggests that our model provides a fair description of the underlying physics: Our self-assembly mechanism is enabled by the presence of a special type of surface states. The formation of surface states is never uniform. Their typically high non-uniformity would imply a significant variance in period length in wires of the same growth run, same diameter, and even within a single wire. Indeed, this variance is commonly observed and has been reported in many papers.

There is, however, another observation in these superlattice nanowires that, at face value, appears to contradict our model. The same periodic structure, that we were able to explain when the periodicity is along the wire, is often observed *across the wire* (Fig. 10). The polar axis "c" of the material is still parallel to the direction of the period, but the wire grows along the non-polar axis "a." This may be straightforwardly explained, if we consider the fact that any growth starts from a nucleus. The nucleus may be viewed as a very short nanowire that develops in the same process proposed in our model. However, further growth takes place in the "a" direction using the nucleus as a template and thus extending the periodic structure sideways.

The same mechanism may also be at work in the case of layers. At the nuclei formation stage, there is practically no difference between a layer and a nanowire. The nuclei develop like small nanowires and then extend laterally keeping the vertical periodicity of the nuclei. However, since the periodicity may fluctuate, the resulting layer should be polycrystalline. Indeed, layers reported in the literature have all been polycrystalline. Small grains with high surface to volume ratio are also compatible with the premise of surface-state-assisted thermal generation that we propose to be the source of electrical charge driving the periodic inversion. Hence, the same model may also explain the formation of superlattice-structured grains in ceramic polycrystalline layers and may also explain why the phenomenon has not been observed in large single crystals.

## VII. CONCLUSION

While much scientific and technological effort is devoted today to bandgap engineering of complicated heterostructures, nature appears to be able to carry out this task seamlessly and effortlessly in the formation of the homologous superlattices. Reviewing the literature, we have seen that intensive efforts to understand the formation of these structures have led to understanding of various subtle details of the structure. However, there has not been a single attempt to explain the formation of the periodic polarity switching which underlies the self-assembly.

In contrast, evidence has been accumulating that the wurtzite structure is predisposed to be affected by extrinsic electric fields because of its natural built-in polar field. To reach from there to the modeling of our self-assembly process requires only the realization of the fact that the main constituent of the homologous superlattices is a wurtzite material, ZnO, having a high spontaneous polarization. Once a strong built-in field is at work, thermal generation of electron-hole pairs can produce charging and an external field to flip over the polarity. Hence, the homologous superlattice self-assembly may not be more than another example, albeit complicated, of the way wurtzite crystals respond to electrostatic phenomena during their growth.

**ACKNOWLEDGEMENTS**

We gratefully acknowledge the support of BSF grant #2015700 and NSF grant # ECCS-1610362.